\documentclass{article}
\usepackage[utf8]{inputenc}
\usepackage{graphicx}
\usepackage{tabularx}
\usepackage{amssymb}
\usepackage{isomath}
\usepackage{siunitx}
\usepackage{amsmath}
\usepackage{amsmath}
\usepackage{amsfonts}
\usepackage{appendix}
\usepackage{fancyhdr}
\begin{document}
	
	
	\begin{titlepage}
		\clearpage\thispagestyle{empty}
		\noindent
		\hrulefill
		\begin{figure}[h!]
			\centering
			\includegraphics[width=2 in]{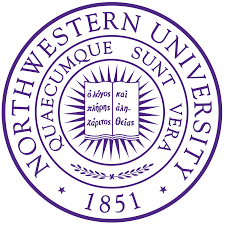}
		\end{figure}
		\begin{center}
			{
				{
					{\bf Center for Sustainable Engineering of Geological and Infrastructure Materials} \\ [0.1in]
					Department of Civil and Environmental Engineering \\ [0.1in]
					McCormick School of Engineering and Applied Science \\ [0.1in]
					Evanston, Illinois 60208, USA
				}
			}
		\end{center} 
		\hrulefill \\ \vskip 2mm
		\vskip 0.5in
		\begin{center}
			{\large {\bf Influence of Geochemistry on Toughening Behavior of Organic-Rich Shale
				}}\\[0.5in]
				{\large {\sc Ange-Therese Akono, Pooyan Kabir}}\\[0.75in]
				{\sf \bf SEGIM INTERNAL REPORT No. 18-3/365I}\\[0.75in]
			\end{center}
			\vskip 5mm
			\noindent {\footnotesize {{\em Submitted to Acta Geotecnica \hfill March 2018} }}
		\end{titlepage}
		
		\newpage
		\clearpage \pagestyle{plain} \setcounter{page}{1}

		\begin{abstract}
Our research objective is to understand the influence of geochemistry on the fracture behavior of organic-rich shale at multiple length-scales. Despite an increasing focus on the fracture behavior of organic-rich shale, the relationships between geochemistry and fracture behavior remain unclear and there is a scarcity of experimental data available. To this end, we carry out 59 mesoscale scratch-based fracture tests on 14 specimens extracted from 7 major gas shale plays both in the United States and in France. Post-scratch testing imaging reveal fractures with small crack width of about 400 nm. The fracture toughness is evaluated using the energetic size effect law, which is extended to generic axisymmetric probes. A nonlinear anisotropic and multiscale fracture behavior is observed. In addition, a positive correlation is found between the fracture toughness and the presence of of kerogen, clay and calcite. Moreover, the geochemistry is found to influence the timescale and the regime of propagation of the hydraulic fracture at the macroscopic length-scale. In particular, shale systems rich in TOC, clay and calcite are more likely to exhibit high values of the fluid lag and low hydraulic crack width. Our findings highlight the need for advanced constitutive models for organic-rich shale systems and advanced hydraulic fracturing solutions that can fully integrate the complex fracture response of organic-rich shale materials.

\end{abstract}

\section{Introduction}
Given the relevance of organic-rich shale in several important energy-related applications such as unconventional gas and oil resources \cite{King2012}, nuclear waste disposal \cite{Lee1991}, and geological sequestration of carbon dioxide \cite{Kang2011,Schepers2009}, a basic understanding of the geomechanical behavior is important. For instance, hydraulic fracturing operations rely on accurate theoretical solutions for the mechanics of hydraulic fracturing \cite{Detournay2016,lecampion2014simultaneous}. In the case of impermeable rocks, the near tip solution and the regimes of propagation are dictated by crucial invariants depending on the local elastic and fracture constants \cite{Detournay2016,WangDetournay2018}. Therefore, a fundamental understanding of the fracture response at the nano- and microscopic length-scales is essential.

Despite advances in elucidating the elasto-plastic response of organic-rich shale systems, the relationships between geochemistry and fracture behavior remain unclear. Although a nonlinear and anistropic fracture behavior has been reported at both the macroscopic \cite{Li2017,Chandler2016} and microscopic scales \cite{Akono2016a}, explanations diverge regarding the dominant fracture mechanisms and the dominant length-scale. On the one hand, the influence of weak bedding planes was emphasized \cite{Chandler2016}, which cannot explain the observed anisotropy at the mesoscale \cite{Kabir2017}. On the other hand, the ductility of kerogen \cite{Abousleiman2016,Hull2017} and the presence of weak kerogen/clay interfaces \cite{Brochard2013} were highlighted. However, the latter explanation implies a positive correlation between the kerogen content or the kerogen+clay content and the fracture resistance. This second hypothesis has not yet been tested due to the scarcity of experimental data available. Thus, the question remains then as to the driving mechanisms---and dominant length-scale---that govern the fracture behavior of organic-rich shale. Therefore, new studies are needed at the granular level or microscopic scale to seek and unravel the links between geochemistry and fracture behavior.

Thus, our research objective is to elucidate the influence of the geochemistry on the fracture behavior by focusing on the granular level. To this end, we examine 10 gas shale materials harvested from 7 major gas shale plays in the United States and in France. Microscopic scratch tests are then utilized to probe the fracture resistance. First, we introduce the materials studied. Then we present our testing procedures: scratch testing and microstructural analysis based on X-ray diffraction and scanning electron microscopy. The theoretical framework for nonlinear fracture mechanics is presented and the size effect law is derived for generic axisymmetric probes. Finally, we explore the correlations between the mechanical and compositional characteristics.

\section{Materials and Methods}
\subsection{Materials}

Our objective is to understand the connection between composition and fracture behavior at the micro- and meso length-scales, as hydraulic fracturing is inherently a multi-scale fracture-driven process. 14 samples were tested in this study, corresponding to 10 gas shale materials that had been extracted from 7 different shale gas reservoirs: Antrim, Fayetteville, Mancos, Marcellus, Niobrara, Toarcian, and Woodford.

\subsection{Microstructural Characterization}
\label{Microstructural Characterization}
In order to yield accurate mechanical measurements at the micro-, and meso- length-scales, it is important to minimize the surface roughness of the polished surface \cite{Miller2008}. The heterogeneous nature of gas shale posed a challenge due to the close intertwining of soft (clay and organic matter) and hard (quartz, feldspar, and pyrite) phases. Grinding was performed using an Ecomet/Automet (Buehler, Lakebluff, IL) grinder polisher in concert with abrasive pads of different gradations. Polishing ensued using either colloidal diamond suspensions, diamond paste, or diamond polishing pads \cite{Akono2016b,Akono2016a}. After grinding and polishing, all specimens were stored under high vacuum to preserve their natural state and prevent potential dehydration or degradation. An important concern was to control the initial saturation level of the organic-rich shale specimens. To this end several routes were pursued. First, great care was taken to avoid exposure to water or water vapor. Second, large specimens were vacuum-sealed in poly-nylon vacuum sealer bags, whereas smaller specimens were stored under 25 Hg vacuum in vacuum dessicators. Finally, testing took place in a hermetically sealed acoustic enclosure.

The mineralogy, composition, and morphology were characterized using X-ray powder diffraction, and TOC coulometer tests, cf. standard ASTM D513. The mineralogy table is given in \ref{Mineralogy}. There is a broad distribution of the clay content ranging from 47.5\% for Fayetteville shale to 1\% for Marcellus shale. Antrim, Fayetteville, Mancos, and Woodford shale systems exhibit a high clay content, whereas Marcellus, Niobrara, and Toarcian exhibit a low clay content. The dominant clay minerals are illite, smectite/illite, and kaolinite. Toarcian and Antrim shale exhibit high quartz content. In contrast, Marcellus, and Niobrara exhibit very low quartz content. Overall, the calcite fraction is very low except for EagleFord, Marcellus, and Niobrara. Meanwhile, the total organic content (TOC) is very high for Woodford, and Antrim shale and the TOC is at its lowest for Toarcian shale specimens. Based on the total organic content, we can classify Marcellus, Niobrara, and Woodford as black shales and Mancos, and Toarcian as gray shales.

\begin{figure} 
	\centering
	\includegraphics[width=\textwidth]{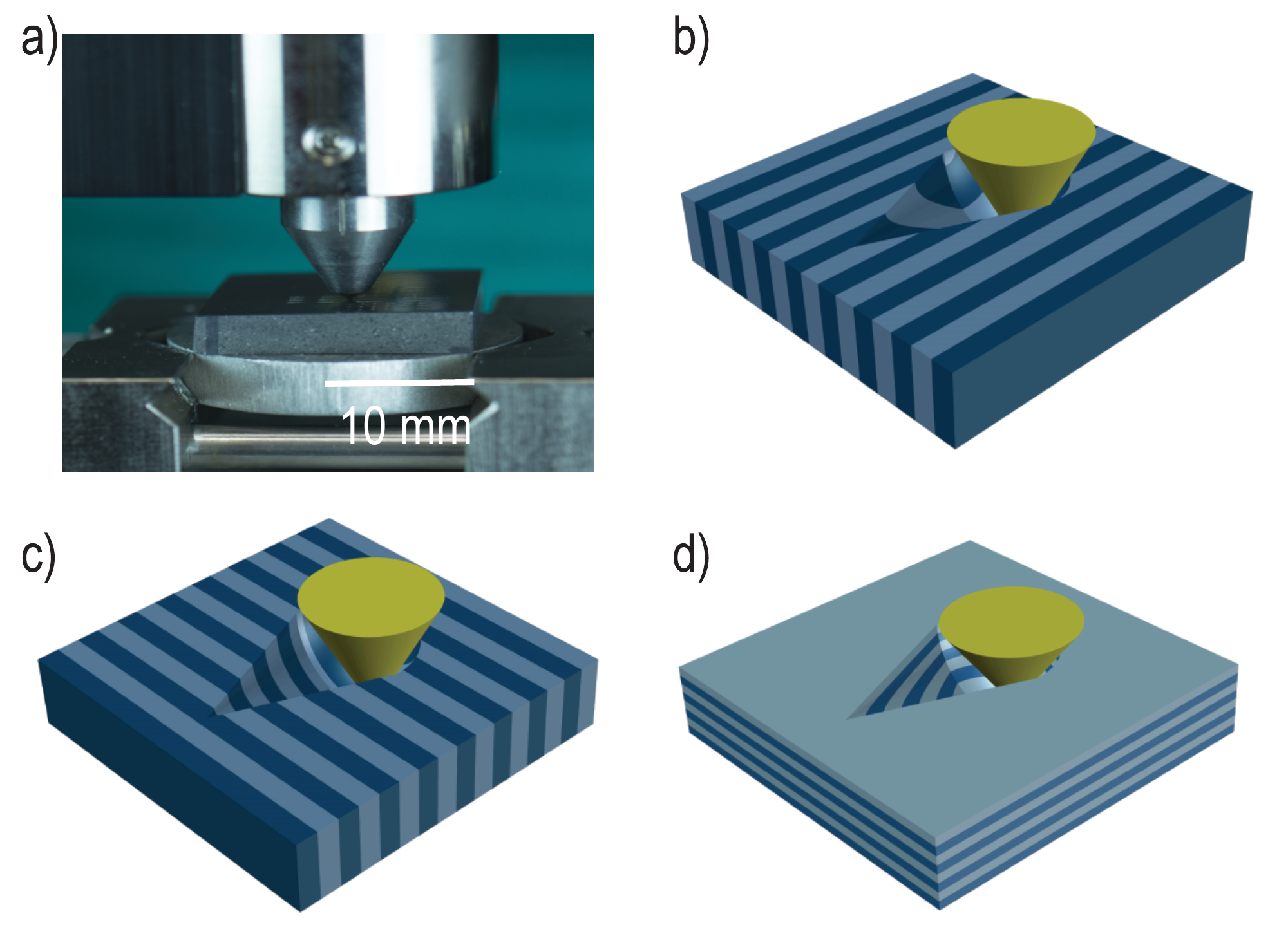}
	\caption{a) Digital photography of a scratch test on a polished Niobrara specimen. Credits: Akono \& Kabir, UIUC, 2016. Scratch orientations: b) divider, c) arrester, and d) short transverse}
	\label{Fig-1}
  \end{figure}

\subsection{Scratch Testing}
Seven years ago, we introduced the method of fracture assessment via scratch testing. The theory and Since then, the method has been widely used to characterize reservoir rocks \cite{2017_Sun_et_al}, tribological coatings \cite{2017_Misra_et_al,2017_Farnoush_et_al}, steel tools \cite{2017_Sola_et_al}, coal \cite{2015_Manjunath_et_al}, and rocks \cite{2017_He_et_al}, and other appications. The scratch test consists of drawing a Rockwell C diamond stylus across the flat surface of the specimen under a progressively increasing vertical load and a constant scratch speed of 6 mm/min. The tests were carried out at a temperature of 22$^\circ$ using an Anton Paar (Anton Paar, Ashland, VA) in an acoustic enclosure. Prior to testing, the stylus was inspected for cracks or damage using an optical microscope and any remaining debris were gently cleaned using an ethanol-saturated cotton swab. Similarly, prior to each test, the surface profile was assessed using a surface profilometer with a contact load of 30 mN. 

A total of 10 gas shale materials, and 14 specimens were tested, with 59 scratch tests carried out. In all tests, the scratch speed was kept constant 6 mm/min, meanwhile the loading rate was 60 N/min for most tests. In all but two tests, the prescribed maximum vertical force was 30 N, whereas the scratch length was 3 mm. Three scratch orientations were considered for Niobrara, and Marcellus: divider, arrester, and short transverse. These three orientations are illustrated in Figure \ref{Fig-1} b--d. In the divider orientation, the crack plane is normal to the bedding plane, and the scratch path moves in a direction parallel to the isotropy plane. In the arrester orientation, both the crack plane and the scratch path direction are normal to the isotropy plane. In the short transverse orientation, the crack plane is parallel to the isotropy plane, and the scratch path moves in a direction parallel to the isotropy plane.

\section{Size Effect Law for Scratch Testing}

Nonlinear fracture mechanics was employed to extract the fracture parameters from the scratch test data. Herein we extend Bazant's size effect law (SEL) to scratch testing with generic axisymmetric probes.
The final equation relating the fracture toughness $K_c$ to the applied force $F_T$ reads \cite{Akono2016jnm,Akono2016a}:
 \begin{equation}
F_{eq}=K_c\sqrt{2pA}
\label{Eq-2}
\end{equation}
where $F_{eq}$ is an equivalent force that accounts for both the vertical and horizontal force along with the stylus back-rake angle. For an inclined stylus, $F_{eq}=\sqrt{F_T^2+\frac{3}{10} F_V^2}$. However, for a straight stylus, $F_{eq}=F_T$.

Starting from Eq.\eqref{Eq-2} and using dimensional analysis, we can write the scratch force as:
\begin{equation}
F_{eq}^2=2pA\frac{K_c^2 }{\mathcal{F}\left(\Pi_1=\frac{l_{ch}}{A/(2p)}, \Pi_2=\frac{l_{ch}}{p},\Pi_3=\frac{H}{M}\right)}
\label{Eq-3}
\end{equation}
Herein, the characteristic length is $l_{ch}=\left(\frac{K_c}{H}\right)^2$ defined as a function of the indentation hardness $H$ to account for the contact pressure \cite{Hubler2016,Hillerborg1976}), $M$ being the indentation modulus. For a conical probe, the quantity $\frac{\Pi_2}{\Pi_1}=\frac{A}{p^2}=\frac{\sin^2\theta}{4\cos\theta}$ depends solely on the probe half-apex angle $\theta$. More generally, for an axisymmetric probe defined by $z=Br^\epsilon$, the quantity $\frac{\Pi_2^\epsilon}{\Pi_1}\approx\frac{2^\epsilon B\epsilon}{\epsilon+1}\frac{B\left(l_{ch}\right)^\epsilon}{l_{ch}}$ is dimensionless and depends essentially on the probe geometrical characteristics---$(B,\epsilon)$---and can be considered a constant with respect to the penetration depth, $d$. We can then write:
\begin{equation}
F_{eq}^2= 2pA\frac{K_c^2}{\mathcal{F}\left(\Pi_1=\frac{l_{ch}}{A/2p}, \tilde{\Pi_2}=\frac{\Pi_2^\epsilon}{\Pi_1},\Pi_3=\frac{H}{M}\right)}
\label{Eq-4}
\end{equation}

For a series of test on a single material---for which $\tilde{\Pi_2}$ is constant---and using the same probe, we can then rewrite:
\begin{equation}
F_{eq}^2=2pA\frac{K_c^2 }{\tilde{\mathcal{F}}\left(\Pi_1=\frac{l_{ch}}{A/2p}\right)}
\label{Eq-4-b}
\end{equation}
Defining the nominal strength by $\sigma_N \equiv \frac{ F_{eq}}{A}$ and the nominal size by $D\equiv \frac{A}{2p}$, we can rewrite: $\sigma_N=\frac{K_c\sqrt{2}}{\sqrt{D\tilde{\mathcal{F}}\left(\frac{l_{ch}}{D}\right)}}$. Thus, we can perform a Taylor expansion of the right hand size of Eq. \eqref{Eq-4-b} with respect to its first argument: $\tilde{\mathcal{F}}\left(\frac{l_{ch}}{D}\right)\approx \tilde{\mathcal{F}}(0)+\tilde{\mathcal{F}}^\prime(0) \frac{l_{ch}}{D}$. We then retrieve the famous energetic size law:
\begin{equation}
\sigma_N=\frac{Bf^\prime_t}{\sqrt{1+\frac{D}{D_0}}}; \quad \sigma_N=\frac{F_{eq}}{A}; \quad D=\frac{A}{2p}
\label{Eq-5}
\end{equation}
with $Bf^\prime_t=\frac{K_c}{\sqrt{\tilde{\mathcal{F}}^\prime(0) l_{ch}}}$ and $D_0=\frac{\tilde{\mathcal{F}}^\prime(0) l_{ch}}{\tilde{\mathcal{F}}(0)}$. This solution for a generic axisymmetric probe is in agreement with the more narrow solutions developed for a prismatic blade \cite{Akono2014efm,Akono2017efm} and a conical probe \cite{Akono2016jnm}. Thus, we have a rigorous framework to calculate the fracture toughness from scratch tests using the energetic size effect law.

In practice, the size effect law parameters, cf. Eq. \eqref{Eq-5}, were computed using a nonlinear constrained optimization scheme. The output of a single scratch test is the load-depth curve, $F_T(d)$, which is recorded every 3 $\mu$m along the scratch path. In order to account for potential manufacturing imperfections, the effective cone angle $\theta$ and sphere radius $R$ are calibrated prior to scratch testing using reference materials \cite{Akono2014wear}. The SEL parameters $Bf^\prime-t$ and $D_0$ are then computed using a truncated Newton algorithm \cite{Nocedal2006} implemented in the programming language Python. We define the $x-y$ variables: $x=\ln(D)$ and $y=\ln(\sigma_N)$ where $\ln$ denotes the natural logarithm function. The model function for the constrained optimization scheme is: $y=\ln\frac{M}{\sqrt{N+\exp(x)}}$. After convergence of the algorithm, the parameters $M$ and $N$ lead to the SEL constant $Bf^\prime_t$ and $D_0$ according to: $D_0=N$, $K_c=M$, $Bf^\prime_t=M/\sqrt{N}$.

\section{Results}

		\begin{figure}
	\centering
	\includegraphics[width=0.7\textwidth]{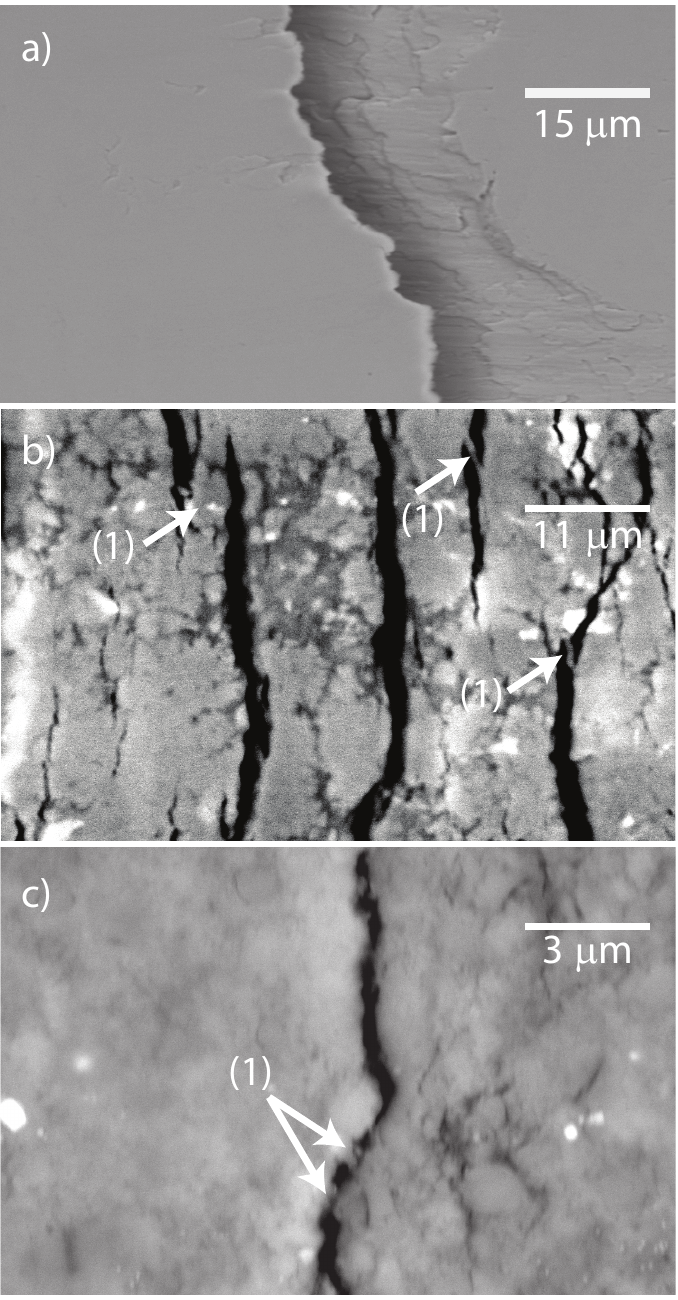}
	\caption{Crack Opening in a) cold-rolled carbon steel, a) silica-enriched cement paste, and c) Niobrara shale. (1) =crack ligament bridging}
	\label{Fig-2}
  \end{figure}

\subsection{Crack Opening in Organic-Shale during Scratch Testing}

Fig. \ref{Fig-2} displays high-resolution scanning electron micrographs of the the residual grooves after scratch testing in cold-rolled carbon steel, silica-enriched cement paste, and Niobrara shale. For all three materials, the scratch length, 3 mm, and scratch speed, 6 mm/min, were the same. Meanwhile the prescribed maximum vertical force was 200 N for cold-rolled carbon steel, and 30 N for silica-enriched cement and Niobrara shale. Fracture surfaces are observable that result from the ductile-to-brittle transition, which is driven by the penetration depth. In cold-rolled carbon steel, which is homogeneous, the crack opening is greater than 18 $\mu$m, whereas for silica-enriched cement, the maximum crack opening is 2 $\mu$m. Finally, for Niobrara shale, the maximum crack opening is 410 nm. 

The differences in crack opening can be explained by the different composition and microstructure along with the different range of fracture micromechanisms in all three materials. In particular, for silica-enriched cement and Niobrara shale, which are both multiscale and highly heterogeneous, we observe crack ligament bridging mechanisms. Furthermore, this small value of the crack opening for organic-rich shale, 411 nm, highlights the multiscale nature of fracture in organic-rich shale within the context of hydraulic fracturing. For instance, at the reservoir scale, lateral fracture as long as 914 m have been reported \cite{2011_Miller_et_al}. In turn, as we will show, the fracture behavior at the macroscopic length-scale is influenced by nano- and microscale compositional features along with a wide range of toughening mechanisms that play out at multiple length scales. 

\begin{figure*}
	\centering
	\includegraphics[width=\textwidth]{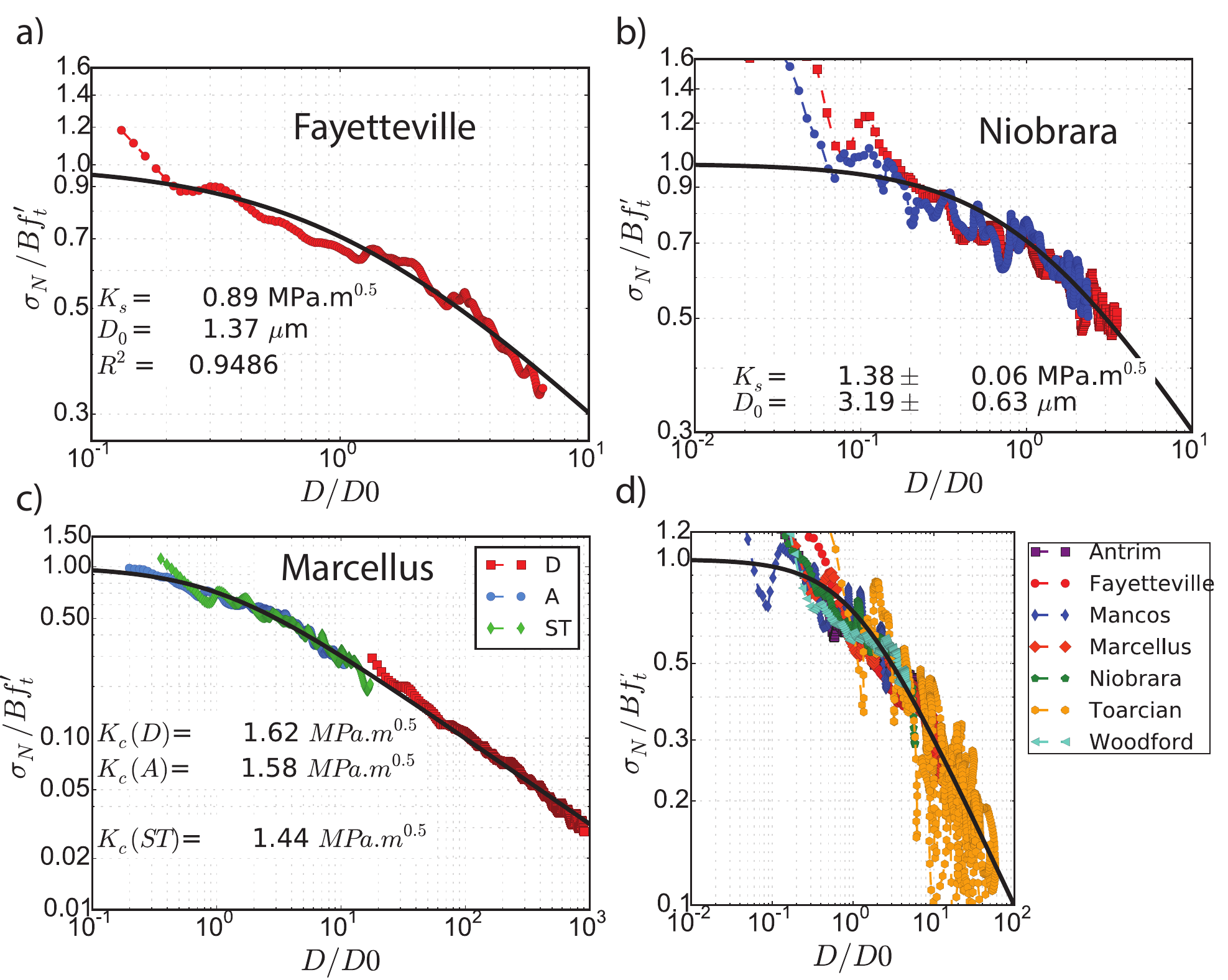}
	\caption{Application of the size effect law, cf. Eq. \eqref{Eq-5}, to scratch tests on organic-rich shale. a) Single tests on Fayetteville shale. b) Three tests on Niobrara shale with the same testing protocol and in the arrester configuration. c) Three tests on Marcellus shale in the divider, arrester and short transverse orientations. d) Compilation of tests on 11 different organic-rich shale materials. $\sigma_N$ is the nominal strength, $D$ is the nominal size, whereas $(Bf^\prime_t, D_0)$ are the size effect law coefficients}
	\label{Fig-3}
  \end{figure*}
	
	\begin{table*}
\centering
\begin{tabularx}{\textwidth}{XXXXX}  \hline 
Material						&	Orientation 			&\# of tests &  $K_c$ (MPa$\sqrt{\mathtt{m}}$)& SEL characteristic length $D_0$ ($\mu$m)  \\ \hline \hline 
Antrim							&	ST	&3&		0.93$\pm$0.076								 &		1.92$\pm$0.34 																	\\
Fayetteville 				&	ST	&3&		0.91$\pm$0.06								 &		0.93$\pm$0.26																	\\
Mancos  						&	ST	&2&		0.97$\pm$0.05								 &		2.47$\pm$0.34																		\\
Marcellus   				&	D		&11&		1.56$\pm$0.28							 &		0.71$\pm$0.44																		\\
Marcellus   				&	A		&11&		1.49$\pm$0.25								&		0.51$\pm$0.83																		\\
Marcellus   				&	ST	&9&		1.43$\pm$0.22								 &		0.42$\pm$1.14																		\\
Niobrara    				&	D   &2&		1.42$\pm$0.22								 &		4.47$\pm$0.60																		\\
Niobrara    				&	A   &3&		1.42$\pm$0.21								 &		3.59$\pm$0.54																		\\
Niobrara    				&	ST	&2&		1.31$\pm$0.16 							 &		1.89$\pm$0.38																		\\
Toarcian B1 &	ST	&3&		0.78$\pm$0.14								 &		0.67$\pm$0.89																		\\
Toarcian B2 				&ST		&2&		0.73$\pm$0.14												 &		0.66$\pm$0.55			        														\\
Toarcian B3 				&	ST	&3&		0.76$\pm$0.15								 &		1.54$\pm$0.52																		\\
Woodford 145 ft 		& ST  &3&		0.62$\pm$0.19								 &		2.42$\pm$0.28																		\\
Woodford 166 ft 		&	ST	&2&		0.63$\pm$0.19								 &		2.40$\pm$0.38																		\\
\hline 
\end{tabularx}
\caption{Summary of the size effect law (SEL) parameters computed for organic-rich shale materials. A=Arrester. D=Divider. ST=Short Transverse. $K_c$ is the fracture toughness whereas $D_0$ is the characteristic length. A total of 59 scratch tests was carried out on 14 gas shale specimens}
\label{Table-4}
\end{table*}

\subsection{Nonlinear Fracture Behavior}
\label{Fracture Scaling}

Fig. \ref{Fig-3} displays the application of the nonlinear fracture mechanics model to microscopic scratch tests on organic-rich shale. We adopt a dimensionless representation of the normalized nominal strength, $\sigma_N/Bf^\prime_t$ , as a function of the normalized nominal size, $D/D_0$ in a log-log space. In Fig. \ref{Fig-3} a) a single test on Fayetteville is modeled. The quality of the fit is evaluated through the coefficient of correlation $R^2$ and the root mean squared error, $RMSE$. The brittleness number $\beta=D/D_0$ ranges from 0.1 to 7. The value of $0.1<\beta<10$ points to nonlinear fracture. In turn, the size effect law can be applied to yield $K_c$ and $D_0$ by application of Eq. \eqref{Eq-5}. The methodology can then be applied to study three different tests carried out on Niobrara shale in the arrester orientation, cf. Fig. \ref{Fig-3} b). For each test, the fracture toughness $K_c$ and the SEL characteristic length $D_0$ are computed separately. For all tests, nonlinear fracture is dominant as the brittleness number $\beta=D/D_0$ ranges from 0.06 to 4. In turn, by considering all three tests in an aggregate manner, we estimate bounds on both $K_c$ and $D_0$. In Fig. \ref{Fig-3} c) the SEL scaling is found to be valid even when considering different orientations for scratch tests performed on Marcellus shale: divider, arrester and, short transverse. As shown in Fig. \ref{Fig-1}, in the divider orientation, the crack plane is perpendicular to the bedding plane, whereas, in the short transverse case, the crack plane moves parallel to the bedding plane. For the arrester and short transverse configurations, the brittleness number ranges from 0.1 to 20. However, for the divider orientation, $\beta=D/D_0$ ranges from 20 to 1000. The range of the brittleness number indicates a transition from nonlinear fracture mechanics to the LEFM regime ($\beta>10$) for both the arrester and short transverse orientations. However, for the divider configuration, LEFM prevails from the very start. In turn, we can see that the fracture behavior is dependent on the orientation. Moreover, the fracture toughness, which is calculated separately for each test, varies as a function of the scratch orientation. Overall, these individual tests illustrate a transition from ductile to brittle fracture occurring in scratch tests.

Fig. \ref{Fig-3} d) highlights the generality of the size effect law to capture the physics of scratch testing in organic-rich shale. Seven tests carried out on different shale materials are shown. Overall, the brittleness number ranges from 0.05 to 70. $\beta=D/D_0<0.1$ is the plastic regime which occurs for very small depths. Ultimately, as the depth is increased, there is a transition into the nonlinear fracture regime and the LEFM regime. Thus, in the case of microscratch testing with an axisymmetric probe, we observe a ductile-to-brittle transition, which is driven by the penetration depth. Therefore, the presence of newly created fracture surfaces on the residual groove, cf. Fig. \ref{Fig-2}, and the strong size effect law scaling corroborate the validity of our fracture mechanics approach to yield the fracture toughness based on scratch tests.

Table \ref{Table-4} provides the Size Effect Law parameters for the scratch tests on all 14 gas shale specimens and 59 tests carried out. The graphs of the size effect law modeling are provided in the Supplementary Information materials. The SEL characteristic length $D_0$ ranges from 0.42 $\mu$m to 4.47 $\mu$. The fracture toughness values range from 0.63 MPa$\mathtt{\sqrt{m}}$ to 1.56 MPa$\mathtt{\sqrt{m}}$, which is in the range of values of fracture toughness values for rocks \cite{Ouchterlony1990,Matsuki1991}. The relative variability of the fracture toughness is 16\% in average. This high variability of the fracture toughness is typical of fracture testing on rocks \cite{schmidt1976fracture,1977_Schmidt}. In adiition, Marcellus and Niobrara shale specimens, exhibit a distinct anisotropy: the fracture toughness is highest in the divider orientation and lowest in the short transverse orientation. In brief, we observe a multiscale, nonlinear and, anisotropic fracture response of organic-rich shale. In what follows, we explore the influence of the orientation and mineralogy on the fracture toughness values.

		\begin{figure*}
	\centering
	\includegraphics[width=\textwidth]{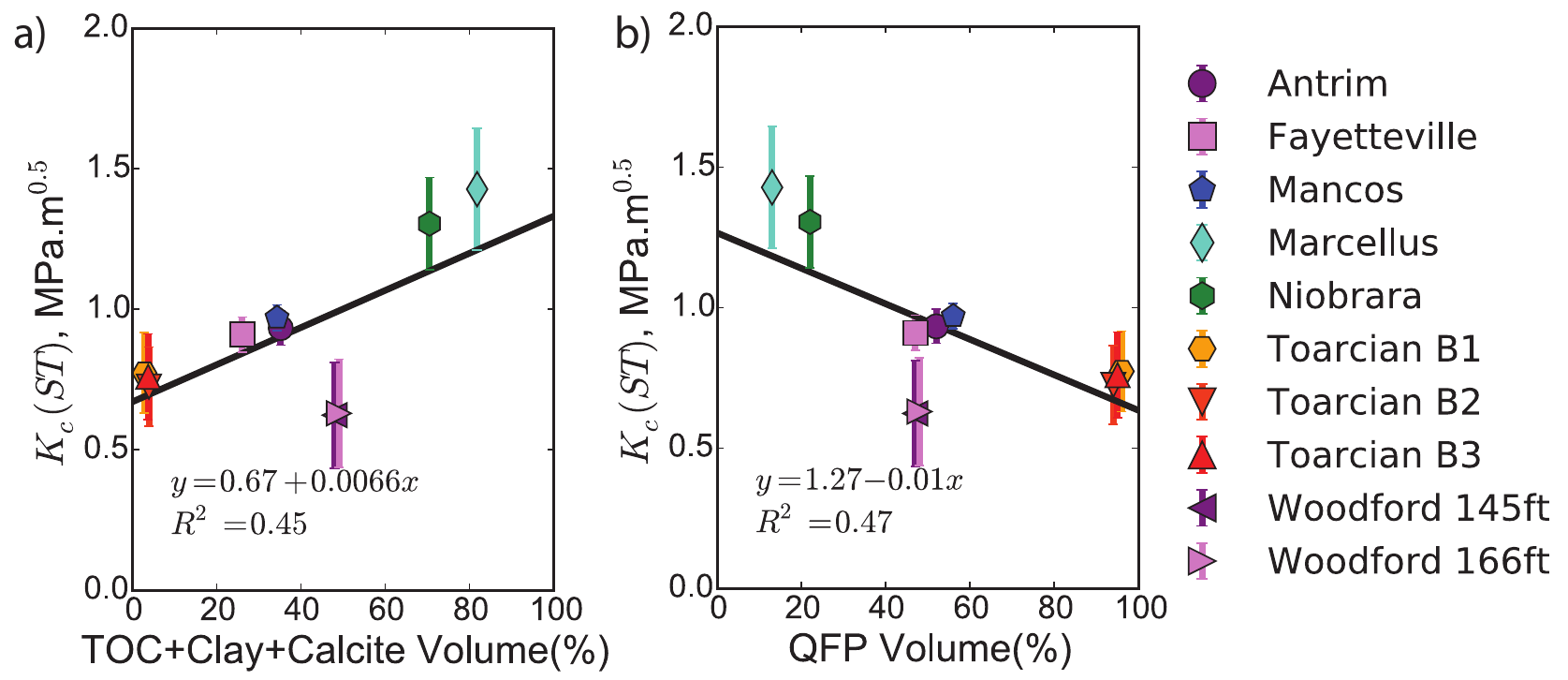}
	\caption{Role of non-clay minerals and organic content of fracture behavior. $K_c(ST)$ is the fracture toughness in the short transverse direction. The graphs show the result of 32 scratch tests. QFP= Quartz, Feldspar, and Pyrite. ST=short transverse}
	\label{Fig-4}
  \end{figure*}
	
\subsection{Impact of Soft Phases: Kerogen, Clay and Calcite}

We  seek to understand the influence of clay and TOC on the fracture resistance. Organic matter is commonly thought to contribute to ductility in organic-rich shale. Clay is thought to contribute to higher values of the fracture resistance via interaction between clay nanoplatelets. 
For instance, \cite{Hull2017} carried out bending tests on organic-rich shale micro-beam and reported a high tensile strength for kerogen. \cite{Sone2013b} reported an increase in ductility with the combined clay and kerogen volume fraction. In our experiments, no conclusive correlation was found between the TOC alone or the clay content alone and the fracture toughness. However, a faint positive correlation was found between the combined amount of TOC, clay and calcite and the fracture toughness.

Fig. \ref{Fig-4} c) displays the fracture toughness in the short transverse orientation versus the combined amount of kerogen, clay, and calcite. The short transverse orientation is selected as it represents a lower bound on the fracture resistance. A positive linear correlation is observed: $y=0.67+0.0066x$ with a coefficient of correlation $R^2=0.45$. This positive correlation suggests that the higher toughness values of kerogen-rich shale is due to the concerted action of total organic content, clay minerals, and calcite. To our knowledge, this is the first study that emphasizes the role of calcite in enhancing the fracture resistance of gas shale systems. In reverse, the fracture toughness is found to linearly decrease as the QFP volume fraction increases, see Fig. \ref{Fig-4} d). In other words, the presence of hard non clay minerals such as quartz, feldspar and pyrite contribute to brittleness.

		\begin{figure*}
	\centering
	\includegraphics[width=\textwidth]{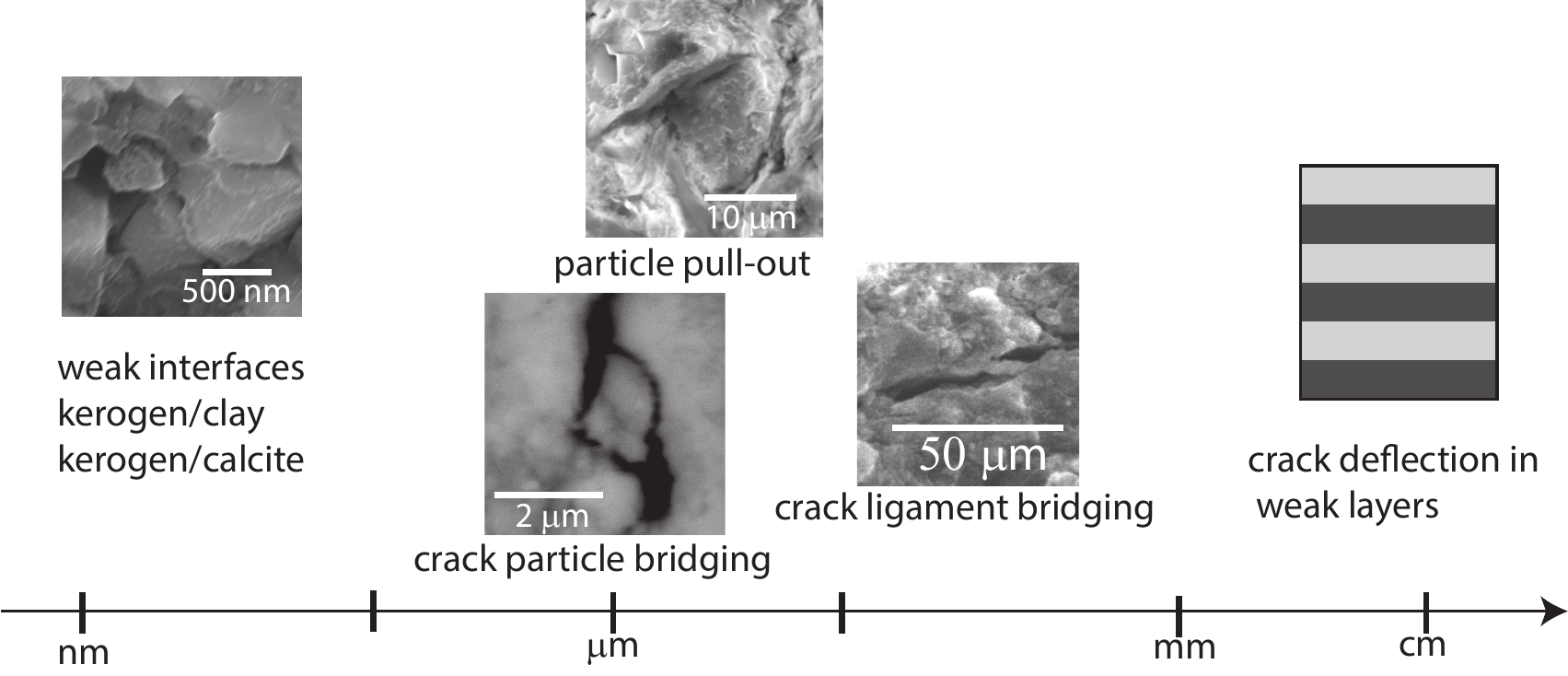}
	\caption{Toughening behavior of organic-rich shale at multiple lengthscales}
	\label{Fig-5}
  \end{figure*}

\section{Discussion}
\subsection{Toughening Mechanisms of Organic-Rich Shale Systems}

Based on our observations of a multiscale nonlinear fracture response, we can draw a map of toughening mechanisms at multiple length-scales. At the nanometer length-scale, to understand the beneficial role of calcite in enhancing the fracture toughness, we turn to biomaterials with a focus on the effect of weak interfaces in natural composites. Weak interfaces are thought to promote crack deflection and crack bridging, leading to a higher roughness of the fracture surface and a higher energy dissipated \cite{Dastjerdi2013,Mirkhalaf2014}. For instance, in nacre, thanks to a three-dimensional brick-wall architecture and weak aragonite/biopolymer interfaces, the crack is channeled around the aragonite tablets, resulting in a three-fold increase in fracture toughness \cite{Barthelat2006,Barthelat2007}. We postulate that a similar intrinsic toughening mechanism is at play for organic-rich shale systems. Weak clay-kerogen and calcite-kerogen interfaces act as sites for crack deflection and crack bridging, leading to an amplification of the fracture toughness. Weak interfaces have been previously identified in organic-rich shale. For instance, using molecular dynamic simulations, \cite{Brochard2013} found out that the kerogen-illite interface is more brittle compared to illite or kerogen separately. However, further investigation is needed with a focus on kerogen-calcite interfaces. Nevertheless, our evidence highlights the important role of calcite, clay and kerogen in enhancing the fracture resistance of gas shale materials.

At the microscopic length-scale, common toughening mechanisms include crack ligament bridging, as shown in Fig. \ref{Fig-2} for Niobrara shale, crack particle bridging \cite{Kabir2017}, distributed microcracking \cite{Kabir2017}, and particle pull-out \cite{Akono2016a,Brochard2013}. At the macroscopic scale, the anisotropy of the fracture behavior was attributed to the layered macro-structure and the presence of weak planes, or calcite veins, which promoted crack deflection \cite{Chandler2016,Li2017}. Nevertheless, our experiments indicate that, even at the granular level, the fracture response is dependent on the orientation. Thus, several mechanisms active at different length-scales contribute to the high fracture toughness values of organic-rich shale systems. In the next section, we evaluate the influence of the geochemistry on the hydraulic fracture solution at the reservoir scale by focusing on five shale systems.

		\begin{figure}
	\centering
	\includegraphics[width=\columnwidth]{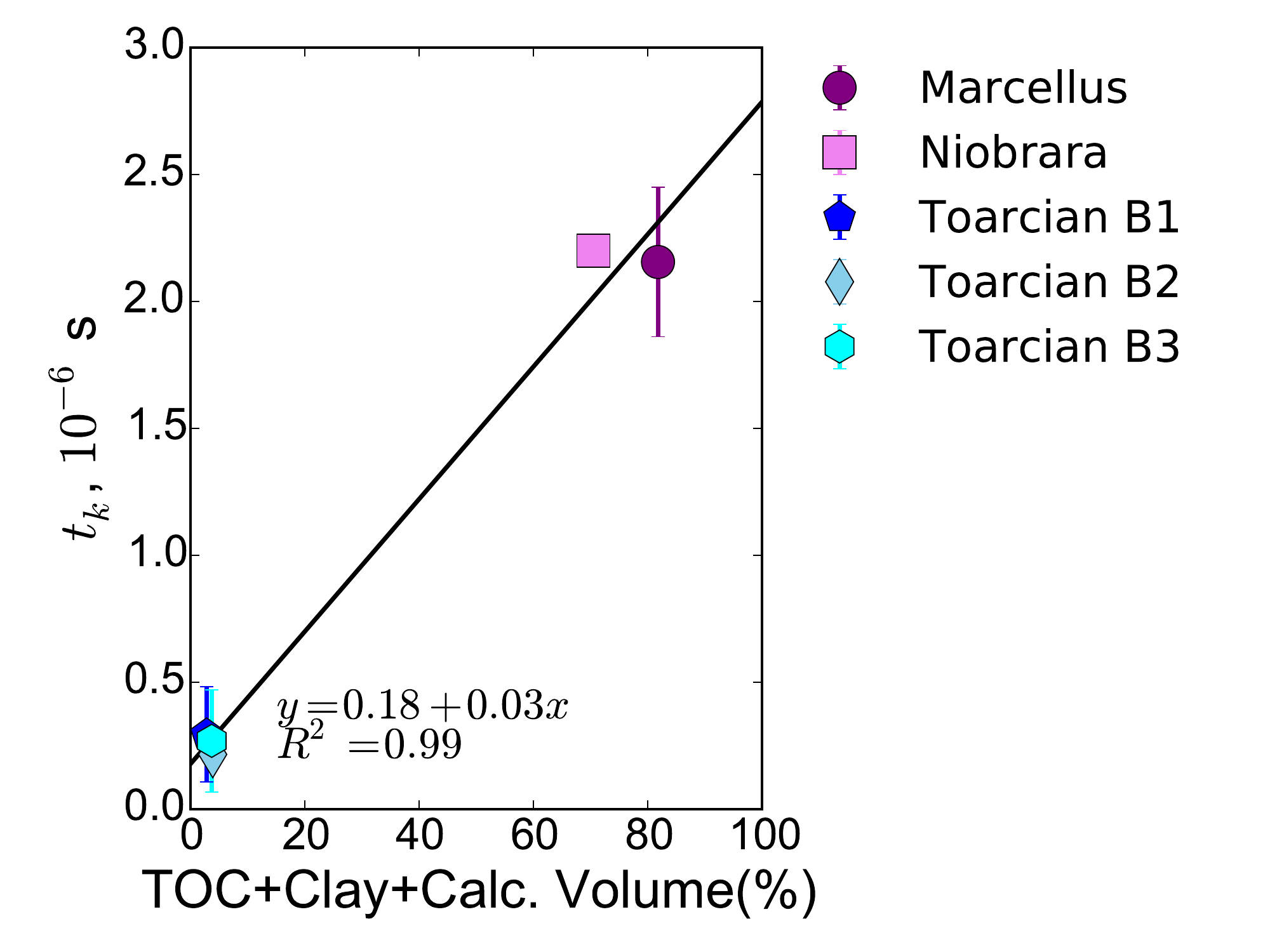}
	\caption{Influence of Geochemistry on timescale of hydraulic fracturing }
	\label{Fig-6}
  \end{figure}
	
		\begin{figure*}
	\centering
	\includegraphics[width=\textwidth]{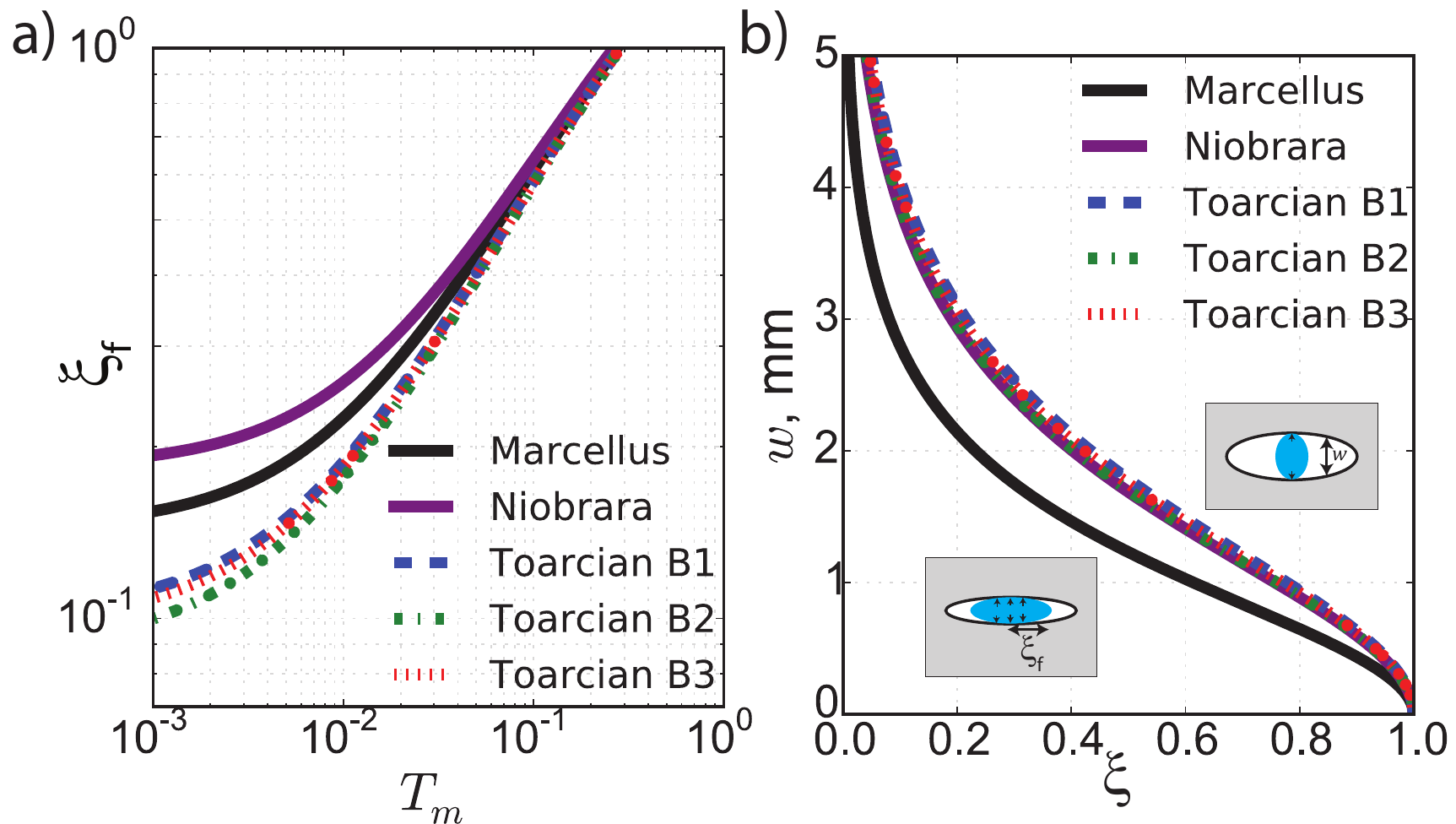}
	\caption{Influence of Geochemistry on Hydraulic Fracturing Solution. a) Fluid lag $\xi_f$. b) Crack opening $w$. $\mathcal{T}_m$ is the underpressure time wehreas $\xi$ is the dimensional coordinate along the hydraulic crack }
	\label{Fig-7}
  \end{figure*}

 \subsection{Influence of Geochemistry on the Early Time Small Underpressure Hydraulic Fracture Solution}

Our objective is to understand the influence of geochemistry on the propagation of the hydraulic fracture at the reservoir length-scale. We rely on existing two-dimensional elastic solutions under plane strain conditions. For organic-rich shale materials that are impermeable, the fracturing fluid leak-off is negligible. Thus we consider an incompressible viscous fluid, of viscosity $\mu$, that is injected at a constant rate $Q_0$ and with a finite non-zero fluid lag, as schematically shown in Fig. \ref{Fig-7} b). The half-crack length is $l$ and the crack opening is $w$. $l_f$ represents the portion of the crack filed by the fracturing fluid and $\xi_f=l_f/l$ is the fluid lag. We consider give shale systems: Marcellus and Niobrara with a high combined fraction of TOC, calcite, and clay, and Toarcian B1, B2, and B3 with low TOC, calcite and clay fractions. On focus is on the evolution of the crack opening $w$ and the fluid lag $\xi_f$ at early times.

The hydraulic fracturing process is defined by important timescale, one of which is the toughness $t_K$ defined by \cite{2006_Garagash}:
\begin{align}
t_k=\frac{M^3}{(-p_t)^3}\frac{(K^\prime)^4}{M^4Q_0}
\label{Eq-6}
\end{align}
where $M$ is the indentation modulus and and $K^\prime=4\sqrt{2/\pi}K_c$. $-p_t\approx \sigma_0$ is the tip underpressure where $\sigma_0$ is the far-field stress, and $Q_0$ is the volumetric injection rate per unit of of out-of-plane fracture width. Fig. \ref{Fig-6} displays the values of $t_k$ as a function of the TOC+Clay+Calc content for the five shale materials considered.For all materials, the toughness timescale is on the order of microseconds. However, a positive correlation is found between the TOC+Clay+Calcite fraction and the toughness scaling. In order words, shale systems rich in organic matter, clay and calcite experience longer toughness scaling.s Thus, the geochemistry influences the timescales of hydraulic fracturing.

The governing equations of the hydraulic fracturing solutions are; lubrication theory, LEFM crack propagation criterion, and fracture opening, provided by the Sneddon and Lowengrubb nonlocal elasticity relation. Herein we adopt the MKO parametric space \cite{Detournay2016,WangDetournay2018,GaragashDetournay2000,SavitskiDetournay2002} where the K-vertex represents the toughness- driven solution whereas the M-vertex represents the viscosity-driven solution. Three invariants are of great importance to capture the physics of the problem \cite{2006_Garagash,Detournay2004}:
\begin{align}
\mathcal{K}_m&=\frac{K^\prime}{M^{3/4}Q_0^{1/4}(\mu^\prime)^{1/4}} \label{Eq-7}\\
\mathcal{M}_k&=\frac{M^3Q_0\mu^\prime}{(K^\prime)^4} \label{Eq-8}\\
\mathcal{T}_m&=\frac{-p_t}{E^\prime}\left(\frac{M t}{\mu^\prime}\right)^{1/3}
\label{Eq-9}
\end{align}
$\mathcal{T}_m$ is the underpressure time, whereas both the viscosity $\mathcal{M}_k$ and  the toughness $\mathcal{K}_m$ scalings determine the propagation regime \cite{2006_Garagash,Detournay2004,Detournay2016}. 

We consider the injection of slick fracking fluid in the impermeable shale formation. Thus we pose: $\mu=0.3$ cP, $\sigma_0=25$ MPa, and $Q_0=0.2$ m$^3$/s. For the materials considered, the toughness scaling $\mathcal{K}_m$ is equal to 0.22, 0.27, 0.17, 0.15, and 0.18,meanwhile, the viscosity scaling $\mathcal{M}_k$ is equal to 438, 201, 1270, 1930, and 1449, respectively for Marcellus, Niobrara, Toarcian B1, Toarcian B2, and Toarcian B3. Herein the indentation modulus $M$ was estimated from 10$\times$10 indentation grids carried out at maximum load of 100 mN. Interestingly, Marcellus and Niobrara, that are rich in TOC, clay and calcite exhibits a higher value of the toughness scaling $\mathcal{K}_m$ compared to Toarcian B1, B2, and B3 systems that exhibit a high fraction of quartz.This observation is consistent with the positive correlation between fracture toughness and TOC+calcite+Clay shown in Fig \ref{Fig-4}. 

Fig. \ref{Fig-7} a) displays the predicted fluid lag $\xi_f$ as a function of the underpressure time $\mathcal{T}_m$ for the five materials of interest. Moreover, Fig. \ref{Fig-7} b) displays the predicted crack opening profile as a function of the dimensional coordinate $\xi=x/l$ along the hydraulic crack. Given the low non-zero values of the toughness scaling and for early propagation times, the small underpressure, small toughness solution prevails, in the O-corner. Thus the fluid lag is solution of an implicit equation \cite{2006_Garagash}. The fluid lag is higher for Marcellus and Niobrara shale and lower for Toarcian shale B1, B2, and B3. These findings suggests that a high TOC+Clay+Calcite content correlates with high values of the fluid lag.

The predicted crack with is displayed, within the small underpressure time and low toughness framework. The early propagation time in Fig. \ref{Fig-6} b) is 1.6 ms. For all five materials, the crack with is in the milimeter range and is maximum at the center. Nevertheless, quart-rich shale materials such as Toarcian B1, B2, and B3 exhibit a larger crack width. In fact, Toarcian shale systems B1, B2, and B3 exhibit a crack width 30\% greater than that of Marcellus shale. The lower predicted carck with for organic-rich shale with a high volume fraction of TOC, Clay and calcite, can be related to the small crack width observed during scratch test experiments in similar systems, cf. Fig. \ref{Fig-2}. Due to toughening mechanisms active at the nano- and microscales---such as crack bridging or crack defelection see map on Fig. \ref{Fig-5}---the crack width does not grow and remains small. In turn, the smaller values of the fracture width will adversely impact the fracture conductivity and thus the efficiency of the hydraulic fracturing process.

Moving forward, our study highlights the need for advanced hydraulic fracturing solution that can fully capture the complexities of the fracture response in organic-rich shale. In this study, a nonlinear, multiscale and anisotropic behavior was observed. Thus, new solutions are needed that integrate the nonlinear fracture behavior, along  with the anisotropy of the fracture toughness, the wide range of toughening mechanisms at different length-scales and the strong fluid-structure coupling. In recent years, some promising approaches have been introduced. A preliminary solution was advanced by \cite{2014_Laubie_Ulm}, however only a linear and weakly coupled system was considered. Similarly a recent study \cite{Zia2018} considered the influence of the anisotropy of the fracture behavior on hydraulic fracturing although the fracture behavior was still linear. Some emerging approaches have been formulated at the mesoscale and macroscale that can incorporate the nonlinear fracture response either based of the lattice particle discrete method \cite{Jin2016} or the microplane models \cite{Lietal2017}. Thus, novel solutions are essential to bridge the scale and fully capture the influence of geochemistry on the hydraulic fracture problem. Nevertheless, the insights provided and the database created will be key to guide and validate these future solutions.

\section{Conclusions}

Our research objective was to elucidate the influence of the geochemistry on the fracture behavior by focusing on the granular level. 14 organic-rich shale materials from major gas shale plays were considered and 59 tests were carried out. In parallel, XRD testing and SEM analysis were employed to characterize the microstructure and mineralogy. The energetic size effect law was derived for scratch testing. When applied to scratch tests on organic-rich shale, the size effect law reveals a transition from ductile behavior at very small penetration depths, to nonlinear fracture and LEFM at large penetration depths. The major findings of our study are summarized below:
\begin{itemize}
\item A small crack opening, ~410 nm, is observed in organic-rich shale.
\item A positive correlation is observed between the combined fraction of TOC, clay and calcite and the fracture toughness.
\item the geochemistry influences the timescale and the regime of propagation of the hydraulic fracture.
\item A high TOC+Clay+Calcite content is correlated with higher values of the fluid lag and lower crack opening at early times of the hydraulic fracture propagation.
\end{itemize}
Thus, our research brings new insights into the fracture response of organic-rich shale and paves the way for advanced physics-based theoretical solutions for hydraulic fracturing in unconventional systems.

\section{Acknowledgments}
This work was supported as part of the Center of Geological Storage of CO$_2$, an Energy Frontier Research Center funded b the U.S. Department of Energy, Office of Science. Data for this project were provided, in part, by work supported by the U.S. Department of Energy under award number DE-FC26-05NT42588 and the Illinois Department of Commerce and Economic Opportunity. The X-ray Diffraction analysis and kerogen content measurements were carried out at the Illinois State Geological Survey XRD Lab. The nanostructural analysis tests was performed in the Frederick Seitz Materials Research Laboratory Central Research Facilities, University of Illinois at Urbana-Champaign. The Woodford shale specimens were provided by the Poromechanics Institute at the University of Oklahoma. The Antrim and Niobrara specimens were provided by the MIT X-Shale project. The Toarcian specimens were provided by the Total Scientific and Technical Center, Pau, France. The Marcellus specimens were provided by the Department of Earth and Planetary Sciences at Northwestern University.

\appendix

\section{Nomenclature}

\begin{table}
\centering
\begin{tabularx}{\textwidth}{XX} \hline 
Symbol 					& Physical Meaning 														\\ \hline
$A$       			& Horizontally projected load-bearing area	 	\\
$B$	  					& Parameter of axisymmetric probe 						\\															
$Bf^\prime_t$ 	& SEL parameter															  \\
$\beta$ 				& SEL brittleness number											\\ 
$\tilde{C}$ 		& Contour for $J$-integral 										\\
$d$       			& Penetration depth 													\\
$D$ 					  & Nominal size															  \\ 
$D_0$ 					& SEL parameter 														  \\ 
$E$			   			& Young's modulus														  \\ 
$\epsilon$		  & Degree of axysimmetric probe 								\\
$F_{eq}$ 				& Effective scratch force 										\\ 
$F_T$     			& Horizontal force														\\ 
$F_V$     			& Vertical force														  \\ 
$\gamma$ 				& Fracture anisotropy ratio 									\\
$\mathcal{G}$ 	& Energy release rate													\\
$H$ 																& Indentation hardness						\\
$K_c$	    													& Fracture toughness 						\\
$\mathcal{K}_m$                     & Toughness scaling               \\
$l_{ch}$ 														& Characteristic length 					\\															
$M$ 																& Indentation modulus 						\\							
$\mathcal{M}_k$                    &   Viscosity scaling           \\
$\mu$                                 & Hydraulic fracturing fluid viscosity\\
$\nu$																& Poisson's ratio							 \\
 $p$       													& Scratch perimeter 						\\
$-p_t$                & Hydraulic crack tip underpressure\\
$Q_0$                  & Hydraulic fracturing constant injection rate\\
 $\psi$ 															& Free energy volume density				 \\
 $R$       													& Probe tip radius							\\
$\underline{\underline{\sigma}}$		& Stress tensor 							\\
$\sigma_N$													& Nominal strength							\\
$\sigma_0$                          & far-field stress              \\
$\theta$   													& Probe half-apex angle						\\	
$\underline{T}$											& Stress vector at stylus-probe interface	\\	
$t_k$                               & Toughness timescale   \\
$\mathcal{T}_m$                       & Underpressure time\\
$\xi$                       & Dimensionless coordinate along hydraulic crack\\
$\xi_f$             & Dimensionless fluid lag\\
$w$                 & Hydraulic crack width \\
\hline 
\end{tabularx}
\caption{Summary and definition of mathematical notations used. SEL=Size Effect Law.}
\label{Table-2}
\end{table}

\section{Mineralogy of Organic-Rich Shale Specimens}
\label{Mineralogy}

\begin{table*}
 \centering
 \begin{tabularx}{\textwidth}{XXXXXXXXX}
 \hline
  Shale        & TOC (\%) & Clay (\%) & Quartz. (\%) & Feld. (\%) & Pyr. (\%) & Calc. (\%) & Other (\%) & Ref\\
 \hline
	Fay. & 4.04  & 22.5  & 47.5  & 0    & 0    & 7.5    & 22.5   & \cite{Bai2013}\\
	Man.       & 1.3  & 24    & 43    & 11   & 2    & 9    & 11     & \cite{Torsaeter2012} \\
	Mar.    & 2.8  & 1     & 10    & 2    & 1    & 78   & 8      &  This Work \\
	Nio.     & 2.5  & 3     & 14    & 4.5  & 3.5  & 65   & 10     & This Work \\
	Toa. B1	 & 0.82 &	2	    & 87	  &8	   &1	    &0     &	2	    & This Work \\
Toa. B2	   & 0.89 &	3	    & 86	    &7	   &1	    &0	   &3	      & This Work \\
Toa. B3	   & 0.71 & 3     &	87	  &7	   &1	    &0.0	   &2	      &  This Work \\
Wf. 145ft &15.4  &27.6	  &34.6	  &4.3	 &7.5   &	0.0	     &10.6	&\cite{Abousleiman2007,Abousleiman2009,Sierra2010} \\
Wf. 166ft &14.6  &	28.2	&32.6	  &4.1   &9.3	  &0.0	    &11.3	& \cite{Abousleiman2007,Abousleiman2009,Sierra2010} \\
 \hline
 \end{tabularx}
\caption{Mineral composition in weight percent of the gas shale specimens tested in this study. TOC=total organic content.
 Calc.=calcite. Feld. = Feldpsar. Pyr.=Pyrite. Other= Dolomite, Siderite, Albite, etc. This Work = specimens tested in this study as described in section \ref{Microstructural Characterization}. Fay.=Fayetteville. Man.=Mancos. Mar.=Marcellus. Nio.=Niobrara. Toa.=Toarcian. Wf.=Woodford.}
 \label{Table-3}
 \end{table*}

\end{document}